\documentclass[amsmath,amssymb,twocolumn]{revtex4-1}
\usepackage{color}
\usepackage{bm}
\usepackage{graphicx}
\usepackage{epsfig}
\usepackage{amsmath}
\usepackage{amsfonts}
\usepackage{amssymb}
\usepackage{bezier}

\def\beq{\begin{equation}}
\def\eeq{\end{equation}}
\def\be{\begin{equation}}
\def\ee{\end{equation}}
\def\bea{\begin{eqnarray}}
\def\eea{\end{eqnarray}}
\def\half{\mbox{$1\over2$}}

\def\bA{{\bf A}}

\def\bk{{\bf k}}

\def\bm{{\bf m}}

\def\bx{{\bf x}}
\def\by{{\bf y}}

\def\bE{{\bf E}}

\def\bd{{\bf d}}

\def\cK{{\cal K}}
\def\cI{{\cal I}}

\def\cA{{\cal A}}

\def\cU{{\cal U}}

\def\cH{{ H}}

\def\cU{{\cal U}}

\def\cS{\mbox{\boldmath $S$}}

\def\cT{{\cal T}}
\def\cV{{\cal V}}

\def\cS{{\cal S}}
\def\cP{{\cal P}}

\def\bD{{\bar{D}}}

\def\hbx{{\hat{\bx}}}
\def\hby{{\hat{\by}}}

\def\hbd{{\hat{\bd}}}

\def\hx{{\hat{x}}}

\def\sd0{\rho^0_s(T)}

\def\1{\mbox{1\hskip-.25em l}}
\def\6{\langle }
\def\9{\rangle }

\def\bsg{{\boldsymbol{\sigma}}}

\def\pk+{\partial_{k+}}

\def\vD{\vec{D}(\bk)}
\def\vtD{\hat{D}(\bk)}
\def\hD{\hat{D}}

\def\Pp{P(\bk)_{+}}
\def\Pm{P(\bk)_{-}}

\def\vg{\vec{\gamma}}

\def\vV{\vec{V}}
\def\vVp{\vec{V_{\perp}}}

\def\Pp1{P_+^{(1)}}
\def\Pm1{P_-^{(1)}}
\def\Ep1{\varepsilon_+^{(1)}(\bk)}
\def\Em1{\varepsilon_-^{(1)}(\bk)}

\def\l{\left}
\def\rr{\right}
\def\hn{{\hat{n}}}
\def\H{{\cal H}}
\def\ket#1{\left|#1\right\rangle}
\def\bra#1{\left\langle #1\right|}
\def\summ{\sum\limits}
\def\bag{\begin{array}}
\def\eag{\end{array}}
\def\be{\begin{equation}}
\def\ee{\end{equation}}
\def\O{{\hat{O}}}
\begin{document}
\parindent=0pt

\title{ Topological Floquet Spectrum in Three Dimensions via a Two-Photon Resonance}

\author{Netanel H. Lindner$^{1,2}$, Doron L. Bergman$^2$, Gil Refael$^2$,
Victor Galitski $^{3,4,5}$}
\affiliation{1) Institute of Quantum Information, California Institute of Technology, Pasadena, CA 91125, USA.}
\affiliation{2) Department of Physics, California Institute of Technology, Pasadena, CA 91125, USA.}
\affiliation{3) Center for Nanophysics and Advanced Materials, Department of Physics,
University of Maryland, College Park, Maryland 20742-4111, USA}
\affiliation{4) Joint Quantum Institute, Department of Physics, University of Maryland, College Park, Maryland 20742, USA.}
\affiliation{5) Kavli Institute for Theoretical Physics, University of California
Santa Barbara, CA 93106-4030}

\begin{abstract}
A recent theoretical work [Nature Phys., \textbf{7}, 490 (2011)]
has demonstrated that external non-equilibrium perturbations may
be used to convert a two-dimensional semiconductor, initially in a
topologically trivial state, into a Floquet topological insulator.
Here, we develop a non-trivial extension of these ideas to
three-dimensional systems. In this case, we show that a two-photon
resonance may provide the necessary twist needed to transform an
initially unremarkable band structure into a topological Floquet
spectrum. We provide both an intuitive, geometrical, picture of
this phenomenon and also support it by an exact solution of a
realistic lattice model that upon irradiation features single
topological Dirac modes at the two-dimensional boundary of the
system. It is shown that the surface spectrum can be controlled by
choosing the polarization and frequency of the driving
electromagnetic field. Specific experimental realizations of a
three-dimensional Floquet topological insulator are proposed.

\end{abstract}
\maketitle
\parindent=15pt

Three dimensional topological insulators exhibit a variety of
novel electronic properties. The most prominent of these are
surface states, whose dispersion is that of massless, chiral two
dimensional Dirac fermions. A dispersion with an odd number of
Dirac cones is unique to surfaces of three dimensional systems, as
it is otherwise precluded by the Fermion doubling theorem. Such
surface states were observed recently by angular resolved emission
spectroscopy in a variety of new materials, such as
$\textrm{Bi}_x\textrm{Sb}_{1-x}$ alloys,
$\textrm{Bi}_2\textrm{Te}_3$,and $\textrm{Bi}_2\textrm{Se}_3$
\cite{Hseih, Xia, Zhang_3D}. The unusual structure of the surface states is predicted to lead to unique response properties of these materials.
Among these is the axion magnetoelectric response \cite{Essin,
Qi_monopole}, which arises when the Dirac cone is gapped due to
breaking of time reversal symmetry, for example by application of
a perpendicular magnetic field. This response is akin to having a
fractional $\nu=1/2$ Hall response at the surface of the material.
Moreover, the surface states of topological insulators play a
crucial role in proposals for quantum interference devices. Most
notable of these is the possibility to realize and manipulate
Majorana Fermions \cite{Fu_majorana}, which have important
applications for topological quantum computing.

The topological behavior of electrons is emerging as a promising
resource, and therefore it is imperative that we understand all
ways to induce it. In this manuscript we explore the possibility
of dynamically inducing a three dimensional topological spectrum,
surface states included, starting with a  trivial
(non-topological) bulk insulator. The idea of inducing topological
order with periodic modulations of a Hamiltonian was explored in
Refs.~\onlinecite{Kitagawa,Kitagawa_walks,Oka,Lindner}. The
Floquet spectrum of a periodically driven system was shown to
exhibit a variety of topological phases, including one that
exhibits a single Dirac cone~\cite{Kitagawa}. Physical examples
include graphene which is expected to exhibit a quantum Hall
effect when subjected to radiation~\cite{Oka, Herb,
demlertransport}, and spin-orbit coupled semiconductor
heterostructure (such as HgTe/CdTe wells), which can be turned
from trivial to topological using microwave-teraHertz
radiation~\cite{Lindner}, and vice versa \cite{Mossner}.

In this manuscript we demonstrate how a ``time reversal
invariant'' three dimensional topological spectrum can be induced
in trivial insulators using electro-magnetic radiation. This
problem is a non-trivial generalization of its 2D
analog~\cite{Lindner}. Roughly speaking, a topologically trivial
band structure can be turned topological by mixing the valence and
conduction bands, for instance by radiative transitions. In 3D,
the radiation has to be carefully tailored such that it produces a
non-vanishing band inversion matrix element in a closed 2D surface
in momentum space, and, as we shall see, must obey additional
topological and symmetry constraints. The polarization and
frequency of the driving electromagnetic field allow for a
detailed engineering of the surface states, including the
possibility to carefully tune a gap in the Dirac cone at the
surface.

First, let us develop our ideas within a simple generic band
structure.  We consider an effective low energy model near the
$\Gamma$ $(\bk=0)$ point
\cite{Zhang_3D}. The four states near the Fermi energy at the $\Gamma$ point are
denoted using two quantum numbers, corresponding to spin $\sigma =
\uparrow,\downarrow$ and parity $\tau=+,-$. Time reversal symmetry is represented by $\cT=i\sigma_y
\cK$, where $\cK$ is the complex conjugation operation. Inversion
symmetry is represented using $\cI= I \otimes \tau_z$. We study a
Hamiltonian of the form
\beq
H= \vec{D}(\bk)\cdot\vec{\gamma}
\label{eq: ham gamma}
\eeq
where $\vg=\left( \gamma_1, \gamma_2, \gamma_3; \gamma_5 \right)$
are four Dirac matrices, which we represent by $ \gamma_i =
\sigma_i \otimes \tau_x$ with  $i=1, 2, 3$, $\gamma_5 = I \otimes
\tau_z$. The remaining dirac matrices are defined as $\gamma_4 = I \otimes
\tau_y$  and  $\gamma_{ij}= -2i\left[\gamma_i,\gamma_j\right]$.
In Eq.~(\ref{eq: ham gamma}) and below, we denote  3-dimensional
(space coordinate) vectors, such as the momentum, $\bk$, in bold
symbols, while the 4-dimensional vectors are denoted with a vector
symbol $\vec{D}$, and $\hat{D}$ for unit vectors. Writing
$\vec{D}(\bk)=\left(\bd(\bk);D_5(\bk)\right)$, we note that the
Hamiltonian~(\ref{eq: ham gamma}) has both time reversal and
space-inversion symmetries under the restriction that the vector
$\bd(\bk)$ be odd under inversion, while $D_5(\bk)$ is an even
function.

Time-reversal-invariant ($\cT^2=-1$ ) band insulators in three
dimensions admit a  $\mathbb{Z}_2$ classification~\cite{Ludwig,
Kitaev, Qi_field}, falling into two categories - either
topological, or trivial. The model in Eq.~(\ref{eq: ham gamma})
can describe both phases, depending on the choice of parameters.
Let us emphasize that while Eq.~(\ref{eq: ham gamma}) does not
describe the most general Hamiltonian with time reversal symmetry
in three dimensions, this effective model does span a wide variety
of realistic systems and allows for a relatively simple
visualization of the $\mathbb{Z}_2$ topological invariant that we
now focus on.

At each momentum, $\bk$, the spectrum of the Hamiltonian
~(\ref{eq: ham gamma}) is doubly degenerate. Its eigenstates
$\psi_\bk$ of~(\ref{eq: ham gamma}) are also the eigenstates of
the rank-two projectors $P_{\pm}(\bk)=\half[I\pm\vtD\cdot \vg]$
onto the valence $(-)$ and conduction bands $(+)$. We can
parameterize the unit vector, $\vtD=\vD/|\vD|$ (lying on a
three-dimensional sphere, $S^3$), using two polar angles, $\theta$
and $\xi$, and an axial angle, $\phi$. We define $\theta$ as
$\cot(\theta_\bk)=D_5(\bk)/|\bd(\bk)|$. The angles $\xi$, $\phi$
correspond to the spin direction, by the unit vector
$\hbd(\xi_\bk,\phi_\bk)=\bd(\bk)/|\bd(\bk)|$. Note that there
exists no global coordinate map on $S^3$ and in our  case,
$\hbd(\xi_\bk,\phi_\bk)$ remains undefined at $\theta=0,\pi$.

Using the above parametrization, the topological invariant for
Hamiltonians of the form~(\ref{eq: ham gamma})  can be calculated
by considering the map from the three dimensional Brillouin zone
(BZ), which is a three dimensional torus, $T^3$, to  $S^3$. The
only topological invariant of this map is an integer which counts
the number of times the map wraps the target space $S^3$, also
called the degree of the map. Two Hamiltonians of the
form~(\ref{eq: ham gamma}), for which the degree of the map
differs by $2$, can be deformed into each other without closing
the gap in the spectrum \cite{NJP}, by adding terms which digress
from the form~(\ref{eq: ham gamma}). Therefore, the $\mathbb{Z}_2$
classification of the insulator~(\ref{eq: ham gamma}) is given by
the degree of the map $\!\!\mod 2$, with an even degree
corresponding to a trivial insulator, and an odd degree to a
topological one.

Near the $\Gamma$ point, an expansion to order $\bk^2$ yields
\beq
\vD=(A\bk; M-B\bk^2),
\label{eq: expansion}
\eeq
where spherical symmetry can be assumed for convenience. The
topological phase of the Hamiltonian~(\ref{eq: ham
gamma}),(\ref{eq: expansion}) occurs for $M/B>0$. In this case,
the angle, $\theta$, changes from $\theta=0$ at $\bk=0$ to
$\theta=\pi$ at $|\bk| \gg
\sqrt {M/B}$. For each $0<\theta<\pi$, the vector $\hbd$ wraps a
sphere of two dimensional unit vectors, $S^2$, which corresponds
to a ``latitude'' on the target space $S^3$. Therefore, the target
space $S^3$ of the map is covered once in the topological phase.

Consider now the case of $M/B<0$. In this phase, the valence band is characterized by  $0\leq\theta<\pi/2$,
while the conduction band corresponds to $\pi/2>\theta\geq\pi$. Therefore, the degree of the map from the BZ to
$S^3$ is zero, leading to a trivial insulator. We note that terms involving other $\gamma$ matrices can be added to $H(\bk)$ in Eq.~(\ref{eq: ham gamma}), while keeping time reversal symmetry. Then,
the $\mathbb{Z}_2$ topological invariant cannot be calculated using the above simple considerations. However, as long as the added terms do not cause the gap in
the spectrum of $H(\bk)$ to close, the $\mathbb{Z}_2$ invariant does not change.

\begin{figure}
\vspace{-1cm}
\includegraphics[scale=0.4]{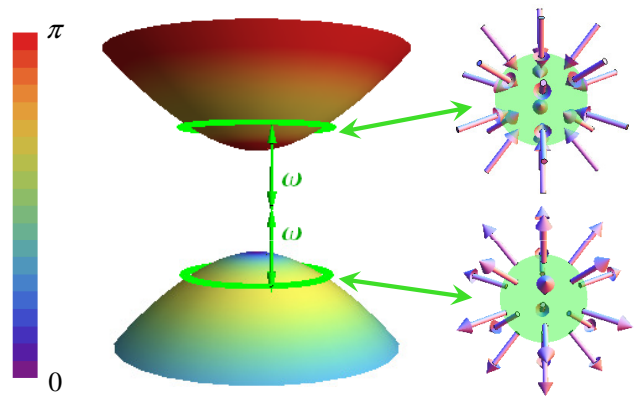}
\vspace{-2cm}
\caption{The two paraboloids represent the dispersion relation $\epsilon_\pm(\bk)$ of the
valence and conduction bands for the Hamiltonian~(\ref{eq: ham
gamma}) in the \textit{trivial} phase, projected on $k_z=0$. Each
energy is doubly degenerate. At each momenta, the eigenstates
of~(\ref{eq: ham gamma}) can be represented by their direction in
$\tau$ space,
$tan(\theta)=|\langle\tau_x\rangle|/\langle\tau_z\rangle$,  and
spin direction $\hbd(\xi,\phi)=(-1)^{{\rm
sign}\langle\tau_x\rangle}\langle\bsg\rangle$ . The color scheme
of the paraboloids represents the value of the angle $\theta$. The
spin direction on spheres in momentum space, $|\bk|=k_0$ is
depicted on the right. A two photon resonance connecting the
valence and conduction bands occurs at such a sphere in momentum
space, and is represented by the green circles on the parabolas.}
\label{fig: dispersion}
\end{figure}

Having reviewed the construction of the  $\mathbb{Z}_2$ topological index within the effective model ~(\ref{eq: ham gamma}), we now discuss a non-equilibrium case,
where a time dependent perturbation is added to Eq.~(\ref{eq: ham gamma}) initially in a trivial phase. As we shall see below, to induce a topological band structure in
three dimensions using electromagnetic radiation, the external field has to satisfy non-trivial constraints. We first illustrate these requirements and elucidate the physics
in the framework of the effective model with a generic time-dependent perturbation of the form:
\beq
V(t)={\rm Re} \left( \vV e^{i\omega t} \right) \cdot \vg
\label{eq: pert}
\eeq
where $\vV$ is a complex, fixed, four-component vector. According
to Floquet theory, the unitary operator describing the
time-evolution of the system, $U(t)=\cP \exp\left(-i \int_{t_0}^t
dt H(t)\right)$, can be written as
\beq
U(t)=W(t)\exp\left[-i H_F(\bk) t\right]
\label{eq: floquet}
\eeq
where $W$ is a unitary matrix satisfying $W(t+T)=W(t)$,
$T=2\pi/\omega$ and $H_F(\bk)$ is a time-independent Floquet
operator. The eigenstates of $H_F$ are solutions of
$\left[-i\partial_t + H(t)\right]\Psi=0$ of the form
$\Psi(t)=e^{-i\varepsilon t}\Phi(t)$, where $\Phi(t)$ is periodic
with the period, $T$, and $\varepsilon$ are called the
quasi-energies. In the following, we shall take the time
independent piece in the Hamiltonian, $H(\bk)$, to be in the
trivial phase, and show that a topological spectrum for the
Floquet operator $H_F(\bk)$ can be achieved nonetheless, by
choosing $V(t)$ appropriately.

In three dimensions, and in the absence of particle-hole or
sublattice symmetries, a topological spectrum requires time
reversal symmetry \cite{Ludwig, Kitaev, Qi_field}. The
instantaneous Hamiltonian $H(\bk,t)=H(\bk)+V(t)$ does not
necessarily posses such symmetry. However, if we can satisfy
\beq
\cT H(\bk,t) \cT^{-1}= H(\bk,-t+\tau)
\label{eq: TRS}
\eeq
for some fixed but arbitrary parameter, $\tau$, then the Floquet operator $H_F(\bk)$ is invariant under an effective time reversal symmetry \cite{Lindner,Kitagawa}. The time-reversal constraint in Eq.~(\ref{eq: TRS}) is satisfied
if $\arg(V_{1,2,3})=\arg(V_{5})+\pi$.

The effect of the time-dependent potential becomes apparent in
rotating frame, where the lower band is shifted up by $\omega$. We
transform the Hamiltonian to such a rotating frame via the unitary
transformation $U(\bk,t)=P_+(\bk)+P_-(\bk)e^{-i\omega t}$ (here
$P_\pm(\bk)$ are the projectors defined above). The resulting
interaction-picture Hamiltonian reads
\bea
H_I(\bk,t)&=&\epsilon_+(\bk)P_{+}(\bk)+\left[\epsilon_-(\bk)
+\omega\right]P_{-}(\bk)\nonumber\\
&+& U(t)V(t)U^{\dag}(t),
\label{eq: HI}
\eea
where $\epsilon_\pm(\bk)=\pm|\vD|$ are the band dispersion
relations of $H(\bk)$. The Floquet operators corresponding to
$H_I(t)$ and to $H(\bk)$ are \textit{identical}, up to a
(time-independent) unitary transformation.

Consider the first two terms in Eq.~(\ref{eq: HI}). In the
interaction picture, the two bands intersect on a two-sphere in
the BZ, $|\bk|=k_0$, if the driving frequency $\omega$ is larger
than the band gap, $2M$. We denote this two-sphere by $\cS$, and
we depict it in Fig~\ref{fig: dispersion}. The driving term (third
term in Eq.~(\ref{eq: HI})) opens a gap in the quasi energy
spectrum at $|\bk|=k_0$.  For momenta away from $\cS$,  the effect
of the driving terms in Eq.~(\ref{eq: HI}) can be neglected
\cite{comment1}, and the quasi-energy states are
roughly eigenstates of $P_\pm(\bk)$. Notice, however, that the
bands of $H_I$ are inverted: the projector $P^I_-(\bk)$ onto the
lower quasi-energy band of $H_I$, corresponds to $P_+(\bk)$ near
the $\Gamma$ point, and to $P_-(\bk)$ for $|\bk|\gg k_0$ (and vice
versa for $P^I_+(\bk)$). The projectors $P^I_\pm(\bk)$ smoothly
interpolate between these two limits.  In order to find the degree
of the map from the BZ to $S^3$ that the projectors $P^I_\pm(\bk)$
define, we need to examine their properties on the resonance
sphere $\cS$.

For values of $\bk$ near the resonance sphere, the rotating wave
approximation (RWA) reveals the effect of the driving field. In
this approximation, only time independent terms are kept in the
last term of Eq.~(\ref{eq: HI}). This yields
\beq
V_{\rm RWA}= \half \Big( P_+ \check{V} P_- + P_- \check{V}^{\dag}.
P_+\Big),\qquad
\check{V}=\vV\cdot\vg
\label{eq: rotV}
\eeq

The topological properties of $H_I$  are intimately related to the
transformation properties of $V_{\rm RWA}$ under the group of
spatial rotations in three dimensions. We shall take the action of
this group in spin space to be $\exp(i\bm\cdot\bsg)$, where $\bm$
parameterizes the axis and angle of rotation. Note that while
$\gamma_5$  transforms trivially under spatial rotations,
$(\gamma_{1},\gamma_2,\gamma_3)$ transform as a vector. Since to
second order in $\bk$ the projectors $P_\pm(\bk)$ are scalars, the
transformation properties of $V_{\rm RWA}$ at this order are
determined by those of $\check{V}$. Therefore, to second order in
$\bk$, $V_{\rm RWA}$ contains both scalar and vector
representations of the rotation group. As we explain below, a
scalar $V_{\rm RWA}$ yields $P^I_\pm(\bk)$ with a non-trivial
topological $\mathbb{Z}_2$ invariant, as it maps the resonance
sphere $\cS$ to cover each spin direction once. However, a purely
vector contribution to $V_{\rm RWA}$ yields a $P^I_\pm(\bk)$ which
is topologically trivial, as the map from $\cS$ covers only a
partial cap of spin-directions.

The explicit form of $V_{\rm RWA}$ is
\beq
V_{\rm RWA} =\half\Big(\vVp\cdot \vg + D_i {\rm
Im}\{V_j\}\gamma_{ij}\Big)
\eeq

where
\beq
\vVp = {\rm Re}\{\vV\}-({\rm
Re}\{\vV\}\cdot\hD)\hD.
\label{eq: rwa1}
\eeq
Two illuminating cases are: (i) $\vV=V_5\hbx_5$ and (ii)
$\vV=V_1\hbx_1$, with $V_1$ and $V_5$ real. This gauge choice
allows us, using Eq.~(\ref{eq: rwa1}), to approximate
$P^I_\pm(\bk)\approx \half (1\pm\hat{n}(\bk)\cdot\gamma)$, and
study the map from the BZ to $S^3$ defined by $\hat{n}(\bk)$.

In case (i), $\check{V}$ is a scalar under spatial rotations, and
hence so is  $V_{\rm RWA}=\vVp\cdot\vg$. Therefore, the first
three components of $\vVp$  must be proportional to $\bk$ while
the fifth components is a scalar. Indeed, using Eq.~(\ref{eq:
rwa1}) we get
\beq
\vVp = \Big(  -\frac{V_5D_5}{D^2}A\bk, V_5-
\frac{V_5D_5^2}{D^2}\Big)
\label{eq: case i}
\eeq
where $D=|\vD|$. On the resonance sphere $\cS$, we have
$\hat{n}(\bk,t)\Big|_\cS=\vVp/|\vVp|$, which maps $\cS$ to an
$S^2$ sphere with fixed latitude $\theta(k_0)$ in the target space
$S^3$. Therefore, $\hat{n}(\bk)$ defines a map from the BZ to
$S^3$, which maps the $\Gamma$ point to the
\textit{south} pole, $S^2$ spheres in the BZ to $S^2$ spheres
(``latitudes'') on $S^3$, and maps $\bk\gg k_0$ towards the
\textit{north} pole of $S^3$.  It is therefore a map of degree one,
which implies a topological spectrum.

In Case (ii), however, $\check{V}$ and  $V_{\rm RWA}$  give a
vector representation of spatial rotations. Therefore, the first
three components of $\vVp$ must either be scalars or belong to a
spin-2 representation of spatial rotations, while the fourth and
fifth components must belong to a vector representation. Indeed,
an explicit calculation gives
\beq
\vVp = \Big( V_1\hbx- V_1\frac{A^2 k_x\bk}{D^2},
V_1\frac{A k_x}{D}\frac{D_5}{D}\Big)
\label{eq: vz v5}
\eeq
Clearly, on $\cS$, the vector
$\vVp$ does not wrap around an $S^2$ sphere on $S^3$, and the
resulting map defined by $\hat{n}(\bk)$ is topologically trivial.

More generally, $V_{\rm RWA}$ will be a superposition of a scalar
and vector component. Whether a topological spectrum is obtained
depends on the relative magnitude of the two components.

Let us now apply the geometrical considerations leading to a
topological Floquet spectrum in 3D, to oscillating electromagnetic
fields. The electric field operator is a vector under spatial
rotations. Therefore, in light of the above discussion, it first
seems impossible that it could induce a topological Floquet
spectrum. This can be remedied by considering multipole tensors of
the electric field. The quadrupole tensor $E_iE_j$ can be
decomposed into a scalar, which is given by its trace, and a
tensor giving a spin-2 representation of spatial rotations. In the
following, we show how to use the scalar part of the quadrupole
tensor in order to induce a topological Floquet spectrum. The
scheme involves choosing a frequency $\omega$ which satisfies
$M/2<\omega<M$. Such a choice for the frequency precludes a
resonance induced by a single-photon transition, which is linear
in the driving electric field $\bE$, but allows for a two-photon
transition, which is second order in $\bE$, and therefore involves
the quadrupole tensor.

Note that in order to satisfy the time reversal symmetry
constraint, as defined by Eq.~(\ref{eq: TRS}), the oscillating field must be
\textit{linearly} polarized \cite{Lindner}. An ellipticity in the
polarization of the driving field leads to interesting effects,
which will be discussed later. We choose a gauge $\bE =
\partial_t
\bA$, $\phi=0$, whereby  the Hamiltonian becomes
$H= \vec{D}(\bk+\bA(t))\cdot\vec{\gamma}$. Choosing $\bA=\cA_0
\cos(\omega t) \hbx$, we obtain the Hamiltonian
\beq
H(t)=\vec{D'}(\bk)\cdot\vg + \vV_1\cdot\vg \cos(\omega t) +
\vV_2\cdot\vg \cos (2\omega t),
\label{eq: hamgaugeB}
\eeq
with $\vV_1 =A\cA_0\hbx-2B\cA_0k_x \hbx_5$, $\vV_2 =-\half
B\cA_0^2 \hbx_5$, and $\vec{D'}(\bk)= \Big(A\bk;
M-B\cA_0^2-B\bk^2\Big)$. Below we sketch the outline of the
calculation, the details of which are given in Appendixes A and B.

The two-photon resonance is a second order process in the electric
field. In the chosen gauge, a $2\omega$-term arises both directly
from $\vV_2$, and from a second order process in $\vV_1$. The
contribution of both terms to $V_{\rm RWA}$ scales, to lowest
order in the light intensity, as $\cA_0^2/M$. In order to
calculate their effect, we perform two consecutive unitary
transformations (see Appendix~\ref{appendix b} for an alternative
derivation). We first perform a transformation to a frame rotating
with frequency $\omega$, of the form leading to Eq.~(\ref{eq:
HI}). The resulting interaction picture Hamiltonian does not
contain any resonances. We diagonalize its time independent terms,
which yields new eigenvalues and projection operators, which we
denote by $\varepsilon^{(1)}_\pm(\bk)$ and $P^{(1)}_\pm(\bk)$
respectively. A second unitary transformation, $U_2(\bk,t)=\Pp1 +
\Pm1
\exp(-i\omega t)$ yields a new interaction picture Hamiltonian
\beq
H_2= \Ep1 \Pp1 + (\Em1+\omega) \Pm1 + U_2\cV^{(1)}(t)U_2^{\dag}
\label{eq: ham2}
\eeq
where \ $\cV^{(1)}(t)$ denotes the time dependent terms resulting
from the first transformation.

After the second transformations, the two bands cross at a surface
$\cS$ with $\Ep1 = (\Em1+\omega)$ and the topology of a sphere. We
now employ the rotating wave approximation, which yields on the
resonance surface,
\beq
H_{2,{\rm RWA}}|_\cS = \half\vec{V}^{(1)}_{\perp} \cdot \vg.
\eeq
The vector $ \vec{V}^{(1)}_{\perp}$ is defined, to lowest order in
$\cA_0$ and $\bk$, as in Eq.~(\ref{eq: rwa1}) with the replacement
$\hat{D}
\rightarrow \hat{D}^{(1)}$ and
$\vV\rightarrow\vec{V}^{(1)}\equiv\left((\vV_1-\vV_2)\cdot\hD\right)\hD+\vV_{2}$
(see Appendix~\ref{appendix a} for details).
The contribution to $\vec{V}^{(1)}_{\perp}$ which allows for a
topological map comes from the term corresponding to $\vV_2$ in
Eq.~(\ref{eq: hamgaugeB}). A quick inspection shows that
$\vV_2\cdot\vg$ is a scalar under spatial rotation, and after the
two transformations yields a contribution to
$\vec{V}^{(1)}_{\perp}$ of the form appearing in Eq.~(\ref{eq:
case i}), with corrections to its spatial components of higher
order in $\bk$ and $\cA_0$. The contribution to
$\vec{V}^{(1)}_{\perp}$ coming from $\vV_{1}$ in Eq.~(\ref{eq:
hamgaugeB}) only leads to an anisotropy of the gap in the Floquet
spectrum, and does not change the topological properties of
$\vec{V}^{(1)}_{\perp}$ on the resonance sphere $\cS$.

\begin{figure}
\includegraphics[scale=0.5]{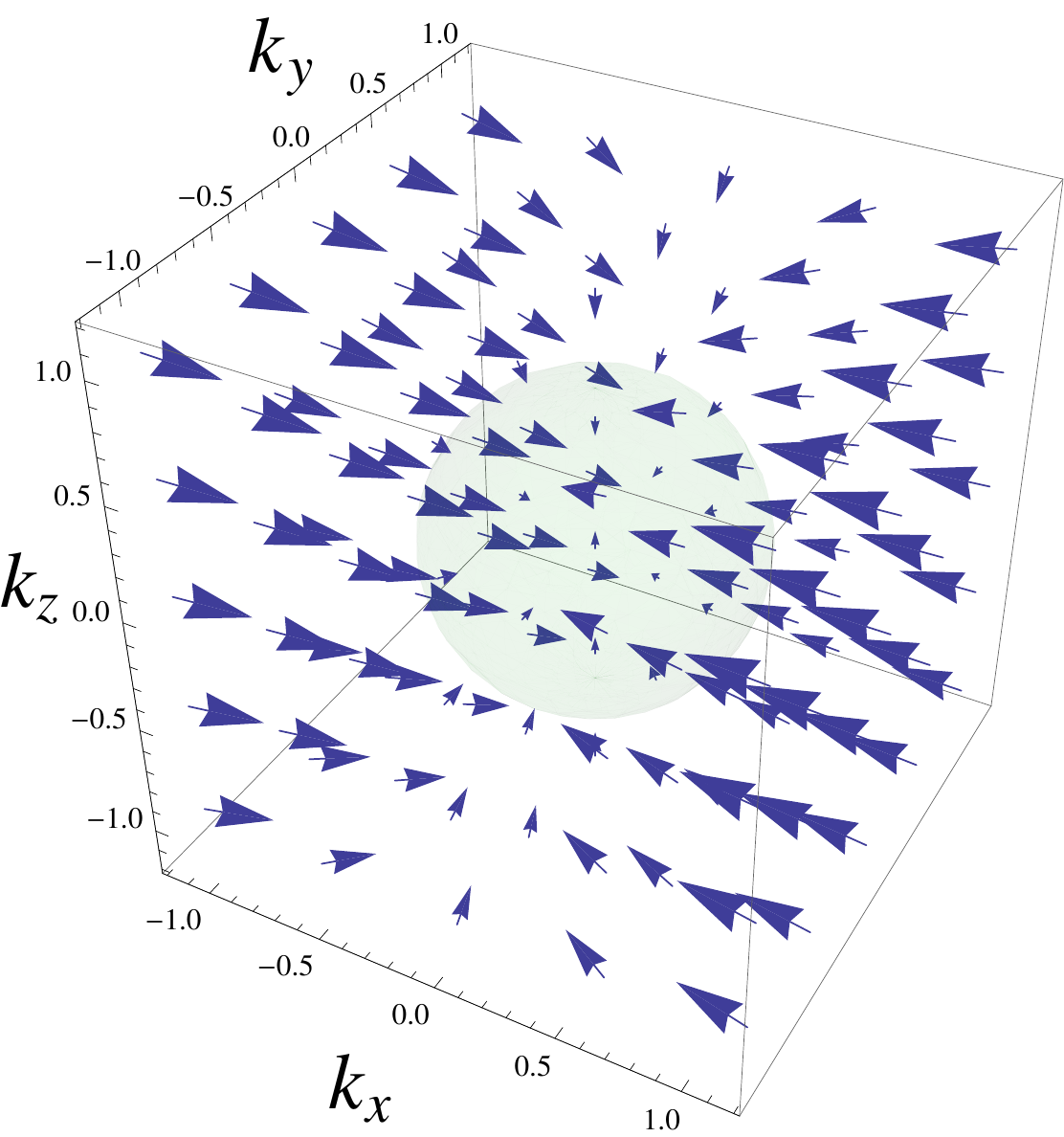}
\caption{
{The first three components of the vector field $\vec{V}^{(1)}_{
\perp}(\bk)$ resulting from a two photon resonance of a linearly polarized electro-magnetic field.
The magnitude of the plotted vector field, on the resonance sphere
$\cS$ (depicted), gives the gap in the Floquet spectrum. Its
direction determines the spin direction of the quasi-energy states
on $\cS$. The map from $\cS$ to the two-sphere $S^2$, defined by
these three components, is of degree one. The resulting gap in the
Floquet spectrum is not isotropic, due to the necessary choice of
the direction of polarization of the electric field. }
}
\label{fig: field}
\end{figure}

The first three components of the vector field
$\vec{V}^{(1)}_{\perp}$ are plotted in Fig.~\ref{fig: field}.
Clearly, they map $\cS$ to a single covering of the unit sphere.
Note that the 5$^{th}$ component of $\vec{V}^{(1)}_{\perp}$ is not
constant on $\cS$. This does not change the degree of the map from
the BZ to $S^3$, and the Floquet spectrum is characterized by a
non-trivial $\mathbb{Z}_2$ invariant. The magnitude of the gap on
the resonance surface is not isotropic. This is expected, as
choosing the polarization direction breaks the rotational symmetry
of the problem. On the resonance, the Floquet spectrum is fully
gapped, where the smallest gap occurs for $k_x=0$, and is given by
\beq
E_{\rm gap}= \frac{A B \cA_0^2}{4M}|\bk_\cS|
\eeq
where $\bk_\cS$ is the resonance wave vector.

%
%

One of the most striking consequences of the topological band
structure in three dimensions are the mid-gap surface modes ,
which are characterized by a single Dirac cone \cite{Fu3D}.
Likewise, a striking consequence of the above considerations are
the appearance of surface modes in the presence of the driving
electric field. The surface quasi-energy states appear inside the
Floquet quasi-energy gap and are characterized by a single Dirac
cone.

\begin{figure}
\vspace{-3cm}
\includegraphics[width=9cm, height=13cm]{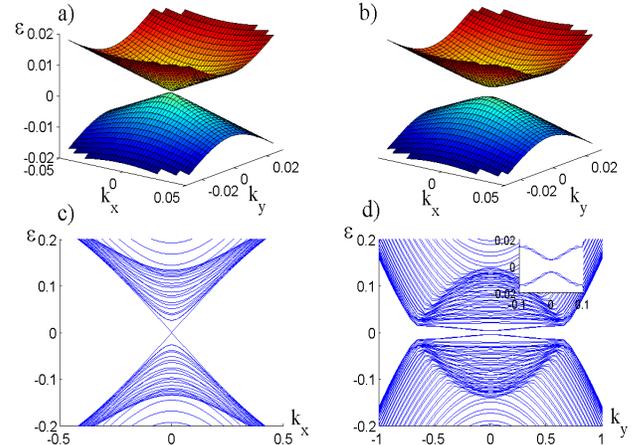}
\vspace{-3.5cm}
\caption{Quasi-energy spectrum of the Floquet operator
corresponding to Eq.~(\ref{eq: hamgaugeB}), in the slab geometry:
periodic boundary conditions in the $x,y$ directions and vanishing
ones in the $z$ direction. In a) and b), we plot the quasi-energy
spectrum inside the gap of the Floquet spectrum, as a function
$k_x,k_y$, for different polarizations of the driving field. a)
Linearly polarized electric field,
$\boldsymbol{\varepsilon}=\hbx$, yielding a Dirac cone on the
surface. b) A small gap in the surface spectrum is opened for an
elliptical polarization with small ellipticity,
$\boldsymbol{\varepsilon}=\hbx+i\delta\hby$, with $\delta=0.05$.
Note that in both a) and b), the Dirac cone is not isotropic. c),
d) Quasi-energy spectrum, showing bulk bands and edge states. c)
Spectrum as function of $k_x$, for $k_y=0$, with
$\boldsymbol{\varepsilon}=\hbx$. d) Quasi-energy spectrum as a
function of $k_y$, for $k_x=0$,
$\boldsymbol{\varepsilon}=\hbx+i\delta\hby$. The inset magnifies
the gapped Dirac surface spectrum. Note that the the gap in the
Floquet spectrum is enhanced for $k_x>0$, as evident by comparing
the spectra in c) and d)}
\label{fig: dirac cone}

\end{figure}

To demonstrate this, we use exact numerical methods to study the
Floquet problem of a tight binding model corresponding to the
Hamiltonian of Eq.~(\ref{eq: ham gamma}). We consider a finite
slab with vanishing boundary conditions at $z=0,L$ and a driving
electric field polarized in the $\hbx$ direction. The quasi energy
and momenta in the $\hbx$ and $\hby$ directions are good quantum
numbers. In Fig.~\ref{fig: dirac cone}, we plot the quasi energy
spectrum inside the quasi-energy gap, which clearly exhibits a
single dirac cone. Note that the Dirac cone is not isotropic,
resulting from the choice of the polarization along the $\hbx$
direction.
%
%

Broken time reversal symmetry on the surface of a three
dimensional topological insulator leads to a gap at the Dirac node
of the surface states. Such a gap entails unique consequences in
terms of transport properties of the surface states, as it leads
to a half integer quantum hall effect on the surface. Remarkably,
a gap at the Dirac node of the surface Floquet spectrum can be
easily controlled by an appropriate tuning of the polarization of
the electromagnetic driving field. Consider an elliptically
polarized electric field, $\bE={\rm Re}
\bE e^{i\omega t}$ with $\boldsymbol{\varepsilon}=\bE/|\bE|=(\hbx+i\delta\hby)$. The
electric field operator is even under time reversal. Therefore
Eq.~(\ref{eq: TRS}) cannot be satisfied for any $\delta\neq0$. The
gap in the Dirac cone on the surface depends quadratically on the
elipticity parameter $\delta$.

The analysis above demonstrates how to obtain a ``time reversal''
invariant topological spectrum in three dimensions, by
periodically driving a topologically trivial system. Our approach
could be applied in  a variety of quantum systems, e.g., cold
atoms with synthetic spin orbit couplings
\cite{galitski,spielman}. The most natural experimental
application of our results are in semiconductors with appropriate
properties, namely, a spectrum with a narrow direct band-gap
occurring in one (or an odd number) of points in the Brillioun
zone. Candidate materials are Sb$_2$Se$_3$  in the rhombohedral
crystal structure
\cite{Zhang_3D}, GeSb$_2$Te$_4$
\cite{Hassan_GbSb2Te4}. Moreover, Heusler compounds \cite{Heusler1, Heusler2}
with applied strain exhibit four bands and a narrow bandgap near
the $\Gamma$ point, and are therefore excellent candidates for our
proposal. In the materials mentioned above, a sizable gap in the
Floquet spectrum, on the order of $10 K$, can be achieved using
electric fields of $10^4 V/m$.

The surface modes in the Floquet spectrum predicted in this work
could be directly probed using photoemission spectroscopy
\cite{Hsieh1,Hsieh2}. Moreover, a gapped surface Floquet spectrum can be
detected using the Kerr and Faraday effect
\cite{Macdonald}, whereas in the material these
effects should be negligible absent the driving. The emergent
Floquet spectrum discussed above is that of a band inverted
semiconductor
\cite{galitski1970}, and in any steady state we can expect a finite density of
particles and holes in the Floquet bands
\cite{glazman},\cite{glazman2}. One concern is that these
free charge carriers may attenuate the driving electromagnetic
field. For a lower bound of the attenuation length $\zeta$ we
assume that all electrons participating in the band inversion act
as free charge carriers. This leads to $\zeta\sim\sqrt{\rho}\sim
1/\sqrt{\mu n}$, with carrier density $n\sim k_0^3$ and mobility
$\mu$. Since the localization length $\lambda$ of the surface
modes scales linearly with $k_0^{-1}$, a parameter regime in which
$\lambda<\xi$ could be found, according to material parameters.

{\em Acknowledgements:} We thank Joseph Avron,  Erez Berg, Daniel
Podolsky, and John Preskill for helpful discussions. VG was
supported by NSF CAREER award. NL was supported by the Gordon and
Betty Moore Foundation and NSF through Caltech's Institute of
Quantum Information and Matter, and by the National Science
Foundation under Grant No. PHY-0803371. DLB was supported by the
Sherman Fairchild foundation. GR and VG acknowledge support from
DARPA. We are also grateful for the hospitality of the Aspen
Physics Center where part of this work was done. We also
acknowledge hospitality of the KITP and the National Science
Foundation under Grant No. NSF PHY05-51164.
\appendix
\section{Topological properties of the two-photon resonance}
\label{appendix a}
In this section we study the topological properties of the Floquet
operator corresponding to an insulator driven with an
electromagnetic field whose frequency allows only for a two photon
resonance, $M<\omega<2M$, where $2M$ is the bandgap of the
insulator. The two-photon resonance is a second order process in
the electric field. In the chosen gauge, a $2\omega$-term arises
both directly from $\vV_2$, and from a second order process in
$\vV_1$. Both terms therefore yield contributions of order
$\cA_0^2$. In order to consider both terms on an equal footing we
shall perform two consecutive time dependent unitary
transformations, where each transformation is characterized by the
frequency $\omega$. In order to analyze the resulting time
dependent Hamiltonian, we shall employ the rotating wave
approximation and expand to lowest orders in $\bk$ and $\cA_0$.

%

%

We first introduce some useful notations. We decompose any four
dimensional vector $\vV$ into the components parallel and
perpendicular to a four dimensional unit vector $\hat{D}$ as
\beq
\vV_{\perp \hD} = \vV-\left(\vV \cdot\hD\right)\hD,
\label{eq: perp}
\eeq
and
\beq
\vV_{\| \hD}=\left(\vV \cdot\hat{D}\right)\hat{D}.
\label{eq: par}
\eeq

For notational convenience, we shall relabel $\vec{D'}(\bk) \to
\vD$, c.f. Eq.~(\ref{eq: hamgaugeB}). The first rotating wave transformation is done via the unitary
\beq
U_1(\bk,t)=P_+(\bk) + P_-(\bk)e^{-i\omega t},
\eeq
which leads to the Hamiltonian in the first rotating frame given
by
\bea
H_1&=& U_1(\bk,t) H(t)  U^{\dag}_1(\bk,t) \nonumber\\
&=& \Big(\vD +\half\vV_{1 \perp \hD}\Big)\cdot\vg+ \omega P_-(\bk)
+
\cV^{(1)}(t).
\label{eq: ham1}
\eea
In the above, the time dependent part is given by
\bea
\cV^{(1)}(t)=&\phantom{+}&(\vV_{1 \| \hD}+\half\vV_{2 \perp \hD})\cdot\vg\cos(\omega t)\nonumber\\
&+&\half(\vV_{2 \perp \hD})_i\hD_j\gamma_{ij}\sin(\omega
t)+\tilde{\cV}(t).
\label{eq: v1t}
\eea
Note that $\cV^{(1)}(t)$ contains terms with frequencies $\omega$
(the first two term in Eq.~(\ref{eq: v1t})) and $2\omega,3\omega$
(corresponding to $\tilde{\cV}(t)$, the third term above).


It is convenient to define
\beq
\hD^{(1)}(\bk)=\Delta\epsilon(\bk)\hD(\bk) +\half\vV_{1\perp\hD}(\bk),
\eeq
with
\beq
\Delta\epsilon(\bk)=\epsilon(\bk)-\half\omega, \qquad
\epsilon(\bk)=|\bar{D}(\bk)|,
\eeq
which enables us to write Eq.~(\ref{eq: ham1}) as
\beq
H_1=\hD^{(1)}(\bk)\cdot\vg + \cV^{(1)}(t)
\eeq

 The time-independent part of Eq.~(\ref{eq: ham1}) can be
expressed using eigenvalues $\varepsilon^{(1)}_\pm(\bk)$ and
projection operators $P^{(1)}_\pm(\bk)$. We now perform a second
rotating wave transformation
\beq
U_2(\bk,t)=\Pp1(\bk) + \Pm1(\bk) \exp(-i\omega t),
\eeq
which yields the Hamiltonian in the $2^{\rm nd}$ frame,
\bea
H_2&=& \Ep1 \Pp1(\bk) + (\Em1+\omega) \Pm1(\bk) \nonumber \\
&+& U_2(\bk,t)\Big((\vV_{1 \| \hD}+\half\vV_{2 \perp
\hD})\cdot\vg \Big) U_2^{\dag}(\bk,t)\cos(\omega t)\nonumber\\ &+&
U_2(\bk,t)\Big((\half\vV_{2 \perp \hD})_i\hD_j\gamma_{ij} \Big)
U_2^{\dag}(\bk,t)\sin(\omega t).
\label{eq: ham2}
\eea
In the above, we have omitted from $H_2$ the term
$U_2\tilde{\cV}(t)U_2^{\dag}$ which does not contain any time
independent contributions to $H_2$, and therefore does not
contribute to the two photon resonance.

After the second transformations, the two bands cross at a surface
$\cS$ for which $\Ep1 = (\Em1+\omega)$. We now employ the rotating
wave approximation. The contribution coming from the second term
in Eq.~(\ref{eq: ham2}) can be deduced from inspecting
Eq.~(8)~and~(9) in the main text. The contribution arising from
the third term in Eq.~(\ref{eq: ham2}) yields a term of the form
$\frac{1}{4i} \hD_i(\vV_{2
\perp \hD})_j\hD_k[\gamma_{ij},\gamma_k]$. Some algebra reveals that to lowest
order in $\cA_0$ and $\bk$, the two terms in Eq.~(\ref{eq: ham2})
involving $\vV_{2\perp \hD}$ yield the same contribution to the
rotating wave approximation.

Therefore, on the surface $\cS$ we have,
\beq
H_{2,{\rm RWA}}|_\cS = \frac{1}{2}\vec{V}^{(1)}_{\perp \hD^{(1)}}
\cdot
\vg
\label{eq: 2photonres}
\eeq
where the vector $ \vec{V}^{(1 )}_{\perp \hD^{(1)}}$ is defined as
in Eq.~(\ref{eq: perp}) by replacing
\beq
\hat{D}
\rightarrow \hat{D}^{(1)}, \qquad \vV\rightarrow\vec{V}^{(1)}=\vV_{1\|\hD}+\vV_{2\perp\hD}.
\eeq

In order to achieve a topological Floquet spectrum, the vector
field $
\vec{V}^{(1)}_{\perp \hD^{(1)}}$ needs to map the
resonance surface $\cS$ to an $S^2$ sphere on the three
dimensional sphere $S^3$. In the following we shall show that this
is indeed the case.

As a first step, we inspect the contributions to $\vec{V}^{(1 )}$,
which arise after the first unitary transformation. Keeping only
terms only up to second order in $\bk$, we find
\beq
 \vV_{1 \|\hD}=\frac{\cA_0 (A^2 - 2 B M)}{M} k_x \Big( \frac{A}{M} \bk,1 \Big), \nonumber\\
\eeq
and
\beq
\vV_{2 \perp\hD}= \frac{A B \cA_0^2}{2M}\left(\bk, -\frac{A}{M} \bk^2
\right).
\label{eq: v2perp calc}
\eeq
The vector field $\vV_{2 \perp}$ clearly maps a sphere in the $BZ$
to an $S^2$ sphere on the target space $S^3$. This is of course
expected from noting that $\vV_2\cdot\vg$ is a scalar under
spatial rotations. However, we are interested in its contribution
to the two-photon resonance,
\textit{i.e.}, to Eq.~(\ref{eq: 2photonres}).

 To this end, we note that
$\vV_{2 \perp\hD}$ is orthogonal to $\hat{D}$ by construction, and
$\hD^{(1)}=\hat{D} +o\Big(\frac{|\vV_{1\perp}|}{\Delta
\epsilon}\Big)\approx\hat{D} +o\Big(\frac{|\cA_0|}{M}\Big)$.
Therefore, the correction to $\vV_{2\perp\hD}$, when it is
inserted into the expression for $
\vec{V}^{(1 )}_{\perp \hD^{(1)}}$, are of higher order in $\cA_0$. Explicitly, we have
\bea
\vV_{2 \perp \hD} \cdot \hat{D}^{(1)} &=& \frac{1}{2|\vec{D}^{(1)}|} \vV_{2 \perp} \cdot
\vV_{1} \nonumber\\
&=& \frac{1}{|\vec{D}^{(1)}|} \frac{A B
\cA_0^3}{4M}\Big(A+2\frac{AB}{M}
\bk^2\Big)k_x\nonumber\\
&\approx&  \frac{1}{\Delta\epsilon(\bk)}\frac{A^2 B
\cA_0^3}{4M}k_x\nonumber
\label{eq: v2perpv1}
\eea
where we have kept terms only up to order $\bk^2$ and $\cA_0^3$.
Therefore, the final contribution of $\vV_2$ to $
\vec{V}^{(1 )}_{\perp \hD^{(1)}}$ to this order is
\beq
\vV_{2 \perp \hD^{(1)}}=\vV_{2 \perp\hD}-\Big(\frac{A^2 B
\cA_0^3}{4M\Delta\epsilon(\bk)}k_x\Big)\hD
\label{eq: v2final}
\eeq

The second term in the above equation correspond to a correction
to the spatial $(1-3)$ parts of $\vV_{2 \perp
\hD}$, Eq.~(\ref{eq: v2perp calc}), of order $\bk^2$ and $\cA_0^3$. The spatial part of
$\vV_{2\perp\hD}$ are originally of order $\bk$ and $\cA_0^2$.
Therefore, to lowest order in $\bk$ and $\cA_0$, this correction
does not alter the topological property of $
\vec{V}^{(1 )}_{\perp \hD^{(1)}}$, which we shall describe
below.

 We now calculate the contribution of $\vV_1$ to $
\vec{V}^{(1 )}_{\perp \hD^{(1)}}$, which will turn out to be of the same order as the contribution of
$\vV_2$, c.f. Eq.~(\ref{eq: v2final}).

First, we note that
\beq
\vV_{1 \|\hD}\cdot \hat{D}^{(1)} =
\frac{\Delta\epsilon}{|\vec{D}^{(1)}|}\left(\vV_{1}\cdot\hD\right)
\label{eq: 2photons}
\eeq

From Eq.~(\ref{eq: 2photons}), and for $\Delta\epsilon(\bk)\gg |
\vV_{1\perp\hD}|$, we have
\bea
\big(\vV_{1 \|\hD}\cdot \hat{D}^{(1)}\big) \hat{D}^{(1)}
&=&\big(1-\frac{|\vV_{1\perp\hD}|^2}{8\Delta\epsilon^2}\big)\left(\vV_{1}\cdot\hD\right)\nonumber\\
&\times&\big(1-\frac{|\vV_{1\perp\hD}|^2}{8\Delta\epsilon^2}\big)\big(\hD+\frac{\vV_{1\perp\hD}}{2\Delta\epsilon}\big)
\label{eq: contr v1 prel}
\eea
Using the definition of $\vV_{1\|}$, Eq.~(\ref{eq: par}), we see
that the total contribution of $\vV_1$ to $
\vec{V}^{(1 )}_{\perp
\hD^{(1)}}$, to order $\cA_0^2$ is
\beq
\vV_{1 \|}-\big(\vV_{1 \|}\cdot \hat{D}^{(1)}\big)
\hat{D}^{(1)}\approx -\frac{\left(\vV_{1}\cdot\hD\right)}{2\Delta\epsilon}\vV_{1\perp}
\label{eq: contr v1}
\eeq
Therefore, this contribution to $
\vec{V}^{(1 )}_{\perp \hD^{(1)}}$ is
of the
\textit{same} order in the driving field as the contribution coming from
$\vV_{2\perp\hD}$, Eq.~(\ref{eq: v2perp calc}) and its inclusion
was necessary for completeness. Note that $\vV_{1\perp}=A\cA_0\hbx
+ o(|\bk|)$, and therefore
\beq
\frac{\left(\vV_{1}\cdot\hD\right)}{2\Delta\epsilon}\vV_{1\perp\hD}\cdot \hbx =\frac{\cA_0^2\big(A^2-2BM\big)}{
2M \Delta\epsilon} A k_x + o(\cA_0^2,|\bk|^3)
\eeq
while the $y$ and $z$ components of Eq.~(\ref{eq: contr v1}) are
of order $\cA_0^2$ and $\bk^3$. Note that the $\hx_5$ component of
Eq.~(\ref{eq: contr v1}) is of order $\cA_0^2$ and $\bk^2$. From
Eq.~(\ref{eq: contr v1 prel}), we see that the $\vV_1$ term also
contributes terms of order $\cA_0^3$ and $\bk^2$ to $\vec{V}^{(1
)}_{\perp
\hD^{(1)}}$. All of the above higher order corrections do not change the topological properties
of $\vec{V}^{(1 )}_{\perp
\hD^{(1)}}$, to lowest order in $\cA_0$ and $\bk$.


Summing up both contributions to $\vec{V}^{(1 )}_{\perp
\hD^{(1)}}$, we have, to lowest order in $\bk$ and $\cA_0$,
\bea
\vec{V}^{(1 )}_{\perp \hD^{(1)}}\cdot \hbx &=&
\Big(\vV_{2\perp\hD}-\frac{\left(\vV_{1}\cdot\hD\right)}{2\Delta\epsilon}\vV_{1\perp}\Big)\cdot
\hbx\nonumber\\
&=& \frac{\cA_0^2}{2M}\Big(B-(A^2-2BM)/\Delta\epsilon\Big)Ak_x,
\label{eq: resx}
\eea
while the other two spatial components are given by
\beq
\vec{V}^{(1 )}_{\perp \hD^{(1)}}\cdot \hbx_\alpha =
\frac{\cA_0^2B}{2M}Ak_\alpha, \qquad \alpha=y,z
\label{eq: resyz}
\eeq

From Eqs.~(\ref{eq: resx},\ref{eq: resyz}), we see that  the
vector field $\vec{V}^{(1 )}_{\perp \hD^{(1)}}$ maps the resonance
surface $\cS$ to an 2-sphere on the target space $S^3$. This
2-sphere is not at a constant ``latitude'', \textit{i.e.} its
$\theta$ coordinate on $S^3$ is not constant. Importantly however,
this 2-sphere winds around the poles of $S^3$,
\textit{i.e.}, it is an incontractible sphere on the space $S^3\setminus(N\cup
S)$, the space of $S^3$ with  the north and south pole removed.


From the above considerations, we see that $H_2(t)$, given in
Eq.~(\ref{eq: ham2}) can be characterized by projection operators
of the form $P^{(2)}_\pm(\bk)=\half(1\pm\hn_2(\bk)\cdot\vg)$. The
unit vector $\hn_2(\bk)$ defines a map from the BZ to $S^3$ with
the properties: (i) For regions in the BZ near the $\Gamma$ point,
$\hn_2(\bk)\approx-\hD(\bk)$, and therefore it maps 2-spheres in
the BZ to 2-spheres on $S^3$, close to its
\textit{south} pole; (ii) Maps the sphere $\cS$ in the BZ to an $S^2$ sphere
 on $S^3$ which is incontractible on $S^3\setminus(N\cup
S)$ (as discussed above) ; (iii) For large values of $\bk$,
$\hn(\bk)\approx\hD(\bk)$,
 therefore these are mapped to 2-spheres  close to the
\textit{north} pole of $S^3$. From continuity of $\hn_2(\bk)$, it
must therefore define a map of degree one. This implies that the
Floquet operator corresponding to $H_2(t)$ and $H(t)$, has a non
trivial $\mathbb{Z}_2$ topological invariant.
\begin{widetext}

\section{Virtual absorption perturbation theory}
\label{appendix b}
The two consecutive RW transformations are very reminiscent of a
perturbation expansion. One RW transformation fails to produce a
degenracy, and therefore a second transformation is necessary to
expose the role of two-photon processes. In this section we will
show how indeed such processes can be analyzed as virtual
absorption processes, and derive a formula which replaces
second-order degenerate perturbation expansions.

The first step involves mapping the time-dependent Floquet problem
to a time-independent problem using an auxiliery degree of
freedom. Let us introduce an additional Hilbert space which serves
as a counter of photons absorbed (for the experts, we note that
this auxiliery variable is just a way of keeping track of the
Floquet block index). We introduce an infinite lattice for a
single particle, which we denote $F$, with states $\ket{n}_F$; n
is essentially counting the number of photons absorbed by the
system.  The original system has a Hamiltonian which is split to
time independent $\H_{sys}$ and time dependent pieces:
\be
H(t)=H_{sys}+\O e^{i\omega t}+\O^{\dagger} e^{-i\omega t},
\label{horg}
\ee
we now replace the time dependent terms with hopping terms for the
register particle F. We also add a diagonal energy term that
determines the energy of the F states. The Hilbert space after
this mapping is a tensor product state between the F-states and
the system's states. The time dependent Hamiltonian is therefore
replaced by an operator that acts on the larger Hilbert space,
\be
\H_{F}=\H_{sys}+\summ_n\l(\O\ket{n+1}_{F~F}\bra{n}+\O^{\dagger}
\ket{n}_{F~F}\bra{n+1}\rr)+\H_\omega,
\label{hf}
\ee
with
\be
\H_{\omega}=\summ_n n\omega \ket{n}_{F~F}\bra{n}.
\ee
and $\H_{sys}=H_{sys}\otimes\mathbb{I}_F$.

To retrieve the original Hamiltonian, Eq. (\ref{horg}) all that is
necessary is to initiate the auxiliary F-states in the
zero-momentum state:
\be
\ket{\psi}_F=\frac{1}{N}\summ_n \ket{n}_F.
\ee
with $N$ providing a normalization.

Our first claim is that the time-independent Hamiltonian, Eq.
(\ref{hf}), initiated with $\ket{\psi}_F$ has the same propagator
for the system as the one for the original Hamiltonian, Eq.
(\ref{horg}):
\be
U(t)=\cP\l[\exp\l[-i\int_{0}^{t} dt
H(t)\rr]\rr]=_F\bra{\psi}\exp[i\H_\omega t]\exp\l[-i
\H_{F} t\rr]\ket{\psi}_F.\label{id1}
\ee
where $\cP$ denotes path ordering. To show this, we first move to
the interaction picture in terms of the states $F$. More precisely
we consider:
\be
\cU(t)=e^{i\H_\omega t}\cdot e^{-i \H_{F}t}.
\ee
We note that
\be
\frac{d \cU(t)}{dt}=-ie^{i \H_{\omega}t}(\H_{F}-\H_\omega)e^{- i \H_{\omega}t} \cU(t).
\ee
Since $\H_{F}-\H_\omega=\H_{sys}+\H_{OF}$, we write:
\be
e^{i \H_{\omega}t}(\H_{F}-\H_\omega)e^{- i
\H_{\omega}t}=\H_{sys}+\H_{OF}(t)
\ee
with:
\be
\H_{OF}(t)=e^{i \H_{\omega}t}\summ_n\l(\O\ket{n+1}_{F~F}\bra{n}+\O^{\dagger}
\ket{n}_{F~F}\bra{n+1}\rr)e^{-i\H_\omega t}=\summ_n\l(\O\ket{n+1}_F~F\bra{n} e^{i\omega t}+\O^{\dagger}
\ket{n}_{F~F}\bra{n+1}e^{-i\omega t}\rr).
\ee
$\cU(t)$ is easily solved to be:
\be
\cU(t)=\cP\l[\exp\l[-i\int^t_0 dt\l( \H_{sys}+\H_{OF}(t)\rr)\rr]\rr]
\ee
And therefore, the identity (\ref{id1}), which we are trying to
prove, becomes
\be
U(t)=_F\bra{\psi} \cU(t)\ket{\psi}_F.
\label{id2}
\ee
Now that we have essentially eliminated $\H_{\omega}$ from the
expression for $U(t)$, the only operators relating to the F-states
remaining are the hopping operators $\summ_n
\ket{n+1}_{F~F}\bra{n}$ and  $\summ_n\ket{n}_{F~F}\bra{n+1}$.
These operators are simple to handle since $\ket{\psi}_F$ is an
eigenstate of both, with eigenvalue 1:
\be
\summ_n\ket{n+1}_{F~F}\bra{n}\psi\rangle_F=\summ_n\ket{n}_{F~F}\bra{n+1}\psi\rangle_F=\ket{\psi}_F.
\ee
Thus we can write:
\be
\H_{OF}(t)\ket{\psi}_F=\ket{\psi}_F \l(\hat{O} e^{i\omega t}+\hat{O}^{\dagger} e^{-i\omega t}\rr)=\ket{\psi}_F H_{O}(t)
\ee
and also:
\bea
\cU(t)\ket{\psi}_F&=&\cP\l[\exp\l[-i\int^t_0 dt\l( \H_{sys}+\H_{OF}(t)\rr)\rr]\rr]\ket{\psi}_F\nonumber\\
&=&\ket{\psi}_F \cP\l[\exp\l[-i\int^t_0 dt\l(
H_{sys}+H_{O}(t)\rr)\rr]\rr]=\ket{\psi}_F U(t)
\eea
which confirms Eq. (\ref{id2}), and therefore completes the proof
of the mapping.

To conclude this section, we note on the correspondence between
the formalism presented above and the Floquet theorem
\beq
U(t)=W(t)\exp\left[-i H_F t\right]
\label{eq: floquet}
\eeq
where $W(t+T)=W(t)$ and $H_F$ is an operator acting on the system
Hilbert space only, see main text, Eq.~(4). The correspondence is
given by noting that choosing $W(t=0)=\mathbb{I}$ gives
\beq
\exp\left[-i H_F t\right]=\bra{\psi}_F\exp\left[-i\H_F t\right]\ket{\psi}_F
\eeq

\subsection{Elimination of single photon processes}

The auxiliery F-states formalism allows accounting for a sequence
of virtual photon absorptions, by systematically eliminating the
F-states associated with intermediate parts of the process. In the
case we considered, for instance, there is no resonant single
photon process. Therefore, if we start the system and F-state wave
function with only an even number of photons, odd-photon F states
will only appear with a suppressed amplitude since they have a
large energy mismatch with the initial states of the wave function
- they must be about an $\omega$ away.

Accounting for two-photon processes in our system is thus possible
along the lines of ordinary second-order perturbation theory. We
start by consdiering the propagator applied to the low-energy
subspace, and read-off the effective hamiltonian that emerges
after resumming connected diagrams. In our case, the low-energy
subspace of the F-states is the superposition of all even states:
\be
\ket{\psi_{even}}_F=\frac{1}{N'}\summ_{n}\ket{2n}_F.
\ee
The effective two-photon propagator is then given by:
\be
U(t)\approx U_{2}(t)=
_F\bra{\psi_{even}}\cU(t)\ket{\psi_{even}}_F.
\label{eq: eff U2}
\ee
Next, we need to expand the interaction Hamiltonian in powers of
$\H_{OF}(t)$.

Before carrying out the expansion, let us move to the interaction
picture of $\H_{sys}$ as well:
\be
e^{i\H_{sys}t}U_{2}(t)\approx _F\bra{\psi_{even}}e^{i\H_{sys}t}
\cU(t)\ket{\psi_{even}}_F=_F\bra{\psi_{even}}\tilde{\cU}(t)\ket{\psi_{even}}_F.
\ee
with
\be
\tilde{\cU}(t)=\cP\l[\exp\l(-i\int_0^t dt \tilde{\cH}_{OF}(t)\rr)\rr].
\ee
We denote
\be
\tilde{\H}_{OF}(t)=e^{i(\H_{sys}+\H_\omega)t} \H_{OF} e^{-i(\H_{sys}+\H_\omega)t}=\summ_n\l(\hat{O}(t)\ket{n+1}_{F~F}\bra{n} e^{i\omega t}+\hat{O}^{\dagger}(t)
\ket{n}_{F~F}\bra{n+1}e^{-i\omega t}\rr).
\ee
with $\O(t)=e^{i\H_{sys}t}\O e^{-i\H_{sys}t}$.

Now we are ready to expand the interaction Hamiltonian in powers
of $\tilde{\H}_{OF}(t)$. Up to second order we encounter the
terms:
\be
\bag{c}
\tilde{\cU}(t)-1=-\int_0^t dt_2 \int^{t_2}_0 dt_1 \l(\O(t_2)
\ket{n+1}_F e^{i(n+1)\omega t_2}+\O^{\dagger}(t_2) \ket{n-1}_F e^{i(n-1)\omega t_2}\rr)\l( _F\bra{n}n\rangle_F e^{-i n\omega (t_2-t_1)}\rr)\\
 \l(\O^{\dagger}(t_1)
_F\bra{n+1} e^{-i(n+1)\omega t_1}+\O(t_1)
_F\bra{n-1}e^{-i(n+1)\omega t_1}\rr).
\eag
\ee
Note that we split the compound: $\ket{n}_{F~F}\bra{m}
e^{-i\omega(m-n) t}=\ket{n}_F e^{i\omega n t}\cdot_F\bra{m}
e^{-i\omega m t}$. The operators $\O,\O^{\dagger}$ could at this
point be construed as {\it first quantized} operators, which
change the state of a particle interacting with the radiation
field.

Further progress is made by projecting on the initial,
intermediate, and final states of the system described by
$\H_{sys}$. Let us denote $P_{\sigma}$ to be a projector of the
system's state on the subspace of energy $\epsilon_{\sigma}$. For
now, we maintain the generality of the discussion, although
eventually, we will restrict ourselves to $H_{sys}=H(\bk)$ which
is a $4\times 4$ Hamiltonian describing two 2d subspaces with
energies $\pm
\epsilon(\bk)$; at that point it will be simple to use $\sigma=\pm 1$
to indicate the valence vs. conduction subspaces. Armed with this
notation we can write:
\be
\bag{c}
\tilde{\cU}(t)-1=-\summ_{\sigma_1,\,\sigma_2,\,\sigma_3}\int_0^t dt_2 \int_{0}^{t_2} dt_1 P_{\sigma_3}\l(\O(t_2)
\ket{n+1}_F e^{i(n+1)\omega t_2}+\O^{\dagger}(t_2) \ket{n-1}_F e^{i(n-1)\omega t_2}\rr)
P_{\sigma_2} e^{-i n \omega (t_2-t_1)}\\
 \l(\O^{\dagger}(t_1)
_F\bra{n+1} e^{-i\omega (n+1)t_1}+\O(t_1)_F\bra{n-1} e^{-i\omega
(n-1)t_1}\rr)P_{\sigma_1}
\eag
\ee
This allows us to resolve the time dependence of the operators:
\be
\bag{c}
\tilde{\cU}(t)-1=-\summ_{\sigma_1,\,\sigma_2,\,\sigma_3}\int_0^t dt_2
\int_{0}^{t_2} dt_1  P_{\sigma_3} e^{-i(\epsilon_{\sigma_2}-\epsilon_{\sigma_3})t_2}\l(\O
\ket{n+1}_F e^{i(n+1)\omega
  t_2}+\O^{\dagger}\ket{n-1}_Fe^{i(n-1)\omega t_2}\rr)
P_{\sigma_2}\\
 \l(\O^{\dagger}
_F\bra{n+1} e^{-i(n+1)\omega t_1}+\O _F\bra{n-1} e^{-i(n-1)\omega
  t_1}\rr)P_{\sigma_1} e^{-i(\epsilon_{\sigma_1}-\epsilon_{\sigma_2})t_1}e^{-i n \omega (t_2-t_1)}
\eag
\ee
The expressions compactify by defining two indices $\mu_{1,2}=\pm
1$, and denoting $\O^{(+1)}=\O^{\dagger}$ and $\O^{(-1)}=\O$, and
dropping  the subscript $F$,
\be
\bag{c}
\Delta
\tilde{\cU}(t)-1=-\summ_{\sigma_1,\,\sigma_2,\,\sigma_3}\summ_{\mu_1,\mu_2=\pm 1} \int^t dt_2
\int_{-\infty}^{t_2} dt_1  P_{\sigma_3} \O^{(-\mu_2)}\\
\ket{n+\mu_2}
P_{\sigma_2} \O^{(\mu_1)} \bra{n+\mu_1}P_{\sigma_1}
e^{-i(\epsilon_{\sigma_1}-\epsilon_{\sigma_2})t_1-i(n+\mu_1)\omega
t_1}
e^{-i(\epsilon_{\sigma_2}-\epsilon_{\sigma_3})t_2+i(n+\mu_2)\omega
t_2} e^{-i n \omega (t_2-t_1)}
\eag
\ee
By moving to average time, $\overline{t}=\frac{t_1+t_2}{2}$ and
time difference, $t_-=t_2-t_1$, as well as integrating over $t_-$
(while assuming that $t$ is large and ignoring boundary terms for
$t_-$, we get:
\be
\tilde{\cU}(t)-1=-\summ_{\sigma_1,\,\sigma_2,\,\sigma_3}\summ_{\mu_1,\mu_2=\pm 1} \int d\overline{t}
 P_{\sigma_3} \O^{(-\mu_2)}
\ket{n+\mu_2}
P_{\sigma_2} \O^{(\mu_1)}  \bra{n+\mu_1}P_{\sigma_1} \frac{ie^{-i
  \overline{t}\l(\epsilon_{\sigma_1}-\epsilon_{\sigma_3}\rr)} e^{-i
  \overline{t}\omega(\mu_1-\mu_2)}}{\frac{\epsilon_{\sigma_1}+\epsilon_{\sigma_3}}{2}+\omega\frac{\mu_1+\mu_2}{2}-\epsilon_{\sigma_2}}
\ee
The time dependence on $\overline{t}$ is simply the
interaction-representation time dependence. Therefore, by going
back to the Schr\"dinger representation, we are able to get rid of
the remaining time dependence in the expression, and we readily
extract the effective second-order contributions to the effective
$\H_F$:
\be
\H^{eff}_{2-ph}=\summ_{\sigma_1,\,\sigma_2,\,\sigma_3}\summ_{\mu_1,\mu_2=\pm 1} \summ_n\ket{n+\mu_2}_{F~F}\bra{n+\mu_1}
\frac{P_{\sigma_3} \O^{(-\mu_2)}
P_{\sigma_2} \O^{(\mu_1)}
P_{\sigma_1}}{\frac{\epsilon_{\sigma_1}+\epsilon_{\sigma_3}}{2}+\omega\frac{\mu_1+\mu_2}{2}-\epsilon_{\sigma_2}},
\label{2ndorder}
\ee
whereby now
\beq
\H^{eff}_F=\H_{sys}+\H^{eff}_{2-ph}+\H_\omega
\eeq
Again we note that $\ket{\psi_{even}}_F$ is an eigenstate with
eigenvalue 1 of $\summ_n\ket{n+\mu_2}_{F~F}\bra{n+\mu_1}$ for
$\mu_{1,2}=\pm 1$, which allows the mapping back to the original
system. As before, the way to go back to the language of the
original time dependent problem is to evaluate Eq.~(\ref{eq: eff
U2}), which yields

\be
U_2(t)=\cP\l[\exp\l[-i\int^t_0 dt\l(
H_{sys}+H^{eff}_{2-ph}(t)\rr)\rr]\rr],
\ee
with
\be
H^{eff}_{2-ph}(t)=\summ_{\sigma_1,\,\sigma_2,\,\sigma_3}\summ_{\mu_1,\mu_2=\pm
  1} \summ_n e^{i(\mu_2-\mu_1)\omega t}
\frac{P_{\sigma_3} \O^{(-\mu_2)}
P_{\sigma_2} \O^{(\mu_1)}
P_{\sigma_1}}{\frac{\epsilon_{\sigma_1}+\epsilon_{\sigma_3}}{2}+\omega\frac{\mu_1+\mu_2}{2}-\epsilon_{\sigma_2}},
\label{2ndorder-2}
\ee

The form of Eq. (\ref{2ndorder-2}) is clearly in accord with
degenerate perturbation theory. The reason for the putative
degeneracy is the fact that we consider the energy of the F states
representing the photons together with the energy of the system. A
resonance, therefore, translates to a degeneracy in this language.
It is interesting to note that Eq. (\ref{2ndorder}) generalizes
degenerate perturbation theory to the case of near degeneracy. The
energy denominator is actually the difference between the average
of the initial and final energies, and the intermediate energy.

\subsection{Application to the 3d FTI construction}

Applying the formalism above to the 3d FTI construction is quite
straightforward. We let
$\O=\O^{\dagger}=\frac{1}{2}\hat{V}=\frac{1}{2}\vec{V}\cdot\vec{\gamma}$,
and
$P_{\sigma}=\frac{1}{2}\l(1+\sigma\frac{H(\bk)}{\epsilon_\bk}\rr)$
with $H(\bk)=\bD\cdot \vec{\gamma}$. We separate to two cases: (1)
$\sigma_1=\sigma_3$, $\mu_1=\mu_2$, and (2)$\sigma_1=-\sigma_3$,
$\mu_1=-\mu_2=-\sigma_1$.

\paragraph{Case 1 - diagonal elements.}  The case of diagonal elements can be treated for both the valence and
conduction band simultaneously, since terms that do not excite
between the bands are time independent, and therefore the f terms
factor out. Therefore:
\be
\H^{eff}_{\sigma_1=\sigma_3}=\ket{n}_{F~F}\bra{n}
\summ_{\sigma_1,\sigma_2=\pm 1,\,\mu=\pm 1}
\frac{P_{\sigma_1}\frac{\hat{V}}{2}P_{\sigma_2}\frac{\hat{V}}{2}P_{\sigma_1}}{(\sigma_1-\sigma_2)\epsilon_\bk-\mu\omega}
\ee
Clearly $\sigma_2=-\sigma_1$, otherwise the denominator makes the
sum vanish. Thus:
\be
\H^{eff}_{\sigma_1=\sigma_3}=\ket{n}_{F~F}\bra{n}
\summ_{\sigma_1=\pm 1}
\frac{P_{\sigma_1}\hat{V}P_{-\sigma_1}\hat{V}P_{\sigma_1}}{4\epsilon_\bk^2-\omega^2}\cdot \sigma_1\epsilon_\bk
\ee
Elementary algebra of the Dirac matrices reduces this expression
to:
\be
\H^{eff}_{\sigma_1=\sigma_3}=\ket{n}_{F~F}\bra{n}\frac{\l(\vec{V}-\frac{1}{\epsilon_\bk^2}\bD\cdot\vec{V}\bD\rr)^2}{4\epsilon_\bk^2-\omega^2}\frac{1}{2}\bD\cdot\vec{\gamma}
\label{eq: diagonal final}
\ee
where we also recognize
$\vV_{\perp\hD}=\vec{V}-\frac{1}{\epsilon_\bk^2}\bD\cdot\vec{V}\bD$.
Note that a term corresponding to Eq.~(\ref{eq: diagonal final})
also arises in the treatment involving the two rotating wave
transformations. Consider the Hamiltonian $H_2(t)$, Eq.~(\ref{eq:
ham2}), evaluated at values of $\bk$ for which
$2\omega=\epsilon(\bk)$, in the rotating wave approximation. For
these $\bk$ values, the terms $\Ep1
\Pp1(\bk) + (\Em1+\omega) \Pm1(\bk)$ do not vanish, but give, to
 order $|\vV_1|^2$ a term corresponding to Eq.(\ref{eq: diagonal
final}).

\paragraph{Case 2 - interband elements.} The interband elements are to
some extent more complicated, since we need to consider excitation
and relaxation separately. We consider a term connecting the
initial state $\sigma_1$ with $\sigma_3=-\sigma_1$. For this
process to be viable, we must have $\mu_1=-\mu_2=-\sigma_1$.
Therefore, the specific term is:
\be
\H^{eff}_{\sigma_1\rightarrow-\sigma_1}=\ket{n+\mu_2}_{F~F}\bra{n+\mu_1}\summ_{\sigma_2=\pm 1}\frac{
  P_{-\sigma_1} \frac{\hat{V}}{2} P_{\sigma_2} \frac{\hat{V}}{2}
  P_{\sigma_1}}{-\sigma\epsilon_\bk}
\ee
Using the form of $P_{\sigma_2}$ we can carry out the sum, and
obtain:
\be
\H^{eff}_{\sigma_1\rightarrow-\sigma_1}=\ket{n+\mu_2}_{F~F}\bra{n+\mu_1}
\frac{  P_{-\sigma_1} \hat{V} H(\bk) \hat{V} P_{\sigma_1}}{-4\epsilon^2_k}
\ee
Once again, elementary manipulations of the Dirac matrices yields:
\be
\H^{eff}_{\sigma_1\rightarrow-\sigma_1}=\frac{1}{4}\ket{n+\mu_2}_{F~F}\bra{n+\mu_1}\l(-\frac{\vec{V}\cdot\bD}{\epsilon_\bk^2}\vec{V}_{\perp\hD}\cdot\vec{\gamma}-\frac{\sigma_1}{2}\frac{\vec{V}\cdot\bD}{\epsilon_\bk^3}\l[\bD\cdot\vec{\gamma},\vec{V}\cdot\vec{\gamma}\rr]\rr).
\ee
This term seems indeed complicated. A simplification occurs,
however, when we consider the RWA directly applied with a
$2\omega$ energy. The effective Hamiltonian connecting the two
bands (that arises from $V$) is (the RWA eliminates the time
dependence, and hence the F's):
\be
\H^{eff}_{interband}=\H^{eff}_{1\rightarrow-1}+\H^{eff}_{-1\rightarrow1}=-\frac{\vec{V}\cdot\hD}{2\epsilon_\bk}\vec{V}_{\perp\hD}\cdot\vec{\gamma}
\label{v1part}
\ee
This term joins the direct two-photon process that arises from the
$\vec{A}^2$ term appearing in $D_5$ due to the mass curvature.
This term also coincides to second order in V, with the
corresponding effect on the radiation-induced  gap within the two
consecutive rotating wave transformations. Note that at resonance
$\epsilon_\bk=2\Delta
\epsilon_\bk$, and thus Eq. (\ref{v1part}) agrees with
Eq. (\ref{eq: resx}).

\end{widetext}

\bibliography{FTI3Drefs}
\end{document}